# Probing electric-dipole-enabled transitions in the excited state of the nitrogen-vacancy center in diamond

Tom Delord[1,*], Richard Monge[1,*], Gabriel Lopez-Morales[1], Olaf Bach[1], Cyrus E. Dreyer[3,4], Johannes Flick[1,2,3], and Carlos A. Meriles[1,2,†]

[1]Department. of Physics, CUNY-City College of New York, New York, NY 10031, USA.
[2]CUNY-Graduate Center, New York, NY 10016, USA.
[3]Center for Computational Quantum Physics, Flatiron Institute, New York, NY 10010, USA.
[4]Department of Physics and Astronomy, Stony Brook University, Stony Brook, New York 11794-3800, USA.

[*]Equally contributing authors
[†]Corresponding author. E-mail: cmeriles@ccny.cuny.edu

The excited orbitals of color centers typically show stronger electric dipoles, which can serve as a resource for entanglement, emission tuning, or electric field sensing. Here, we use resonant laser excitation to expose strong transition dipoles in the excited state (ES) orbitals of the negatively charged nitrogen vacancy center in diamond. By applying microwave electric fields, we perform strong Rabi driving between ES orbitals, and show that the dressed states can be tuned in frequency and are protected against fluctuations of the transverse electric field. In contrast with previous results, we observe sharp microwave resonances between magnetic states of the ES orbitals, and find that they are broadened due to simultaneous electric dipole driving.

Color centers in semiconductors and their electron and nuclear spins are widely used quantum sensors[1] and have attractive low temperature properties for quantum computing[2], but the short range of their magnetic interaction[3] means scalable entanglement between centers needs to be mediated, e.g., by photons[4,5]. In that context, larger electric dipoles in the excited states[6,7] (ES) can be used as an alternative resource for entanglement, resonance tuning[8], or electric sensing[9,10]. The high sensitivity of the ES energies to the solid-state environment, however, also creates challenges in the form of spectral diffusion[6,9–11], screening of the low frequency electric field[12,13], and local heterogeneities[14,15].

Here, we use photo-luminescence excitation (PLE) spectroscopy to study the impact of microwave (mw) electric fields on the excited orbitals of the negatively charged nitrogen-vacancy (NV) center. We first expose a previously unobserved transition dipole moment between the two non-magnetic excited states $E_y$ and $E_x$. PLE spectra reveal the complex features of the dressed state, with the $E_y$ and $E_x$ orbitals interacting with the mw electric field via both their transition and permanent dipole moments. For NVs with moderate strain (in the GHz range), resonant mw fields drive Rabi oscillations in the ultra-strong regime without the need for a resonator. The Rabi splitting we obtain allows us to tune the optical resonances by up to a GHz, while creating new eigenstates protected against the electric noise in the NV transverse plane, effectively reducing the inhomogeneous linewidth by a factor of 1.6. We show that under off-resonance excitation, PLE peaks become flanked by multiple sidebands, their amplitude governed by an interplay between the Stark effect and the driving of the dipole transition. In contrast with previous results, we lastly observe transitions between magnetic and non-magnetic excited states at low temperature[16-18]. Interestingly, we determine that the electric driving broadens the magnetic resonances as it creates multiple peaks in the energy spectrum. While a similar physics is obtained using an acoustic[19] or mechanical[20] resonator to drive the same transitions, an electric drive can be readily tuned on- and off-resonance, and applied through a standard mw antenna. All in all, these results portend fast electrical control of NV ES orbitals without the need for ultra-fast optical pulses[21], which could be applied to electric entanglement of proximal NVs, or to study other electric-active defects in bulk diamond and diamond interfaces.

Our experiments take place at 7K in a closed cycle cryostat, using a home-made confocal microscope to isolate single NVs a few microns deep in a 1-ppm-nitrogen bulk diamond[9,15,22-27]. Importantly, our antenna was not optimized for electric fields: We use a 25 μm wire laid on the diamond crystal to generate both magnetic and electric mw fields. Figure 1a shows a standard PLE spectrum and the measurement sequence we used. A 1 μs green laser pulse initializes the NV into its negative ($m_s = 0$) charge (spin) state before we read the photoluminescence (PL) under excitation by a narrow-band tunable laser resonant with the zero-phonon-line. Applying a pulsed mw magnetic field before read-out allows us to change the spin state and observe the spin dependence of the spectrum. PLE peaks appear when the frequency of the red laser matches one of the six allowed transitions, as depicted in Fig. 1b. Due to the $p$-like character of the ES orbitals, their energies depend on the local strain: In particular, the energies of the non-magnetic orbitals $E_x$ and $E_y$ split (shift) by an energy proportional to the transverse (longitudinal) strain. Throughout this manuscript, we work with an NV of moderate transverse strain, $\delta_\perp = 2.9$ GHz. Figure 1c shows a less conventional PLE measurement where a mw drive runs continuously during optical illumination and PL readout. In this instance, the drive frequency, $\omega_d = 2\pi \times 1.5$ GHz, is far from the ground state magnetic resonance, and we center the PLE spectrum around the $E_x$ and $E_y$ lines, both of which correspond to $m_S = 0$. Surprisingly, the mw drive has a dramatic impact, shifting the main resonances and creating numerous side-peaks at multiples of the drive frequency. We show below that these complex features stem from the combined effects of the transition and permanent electric dipole moments.

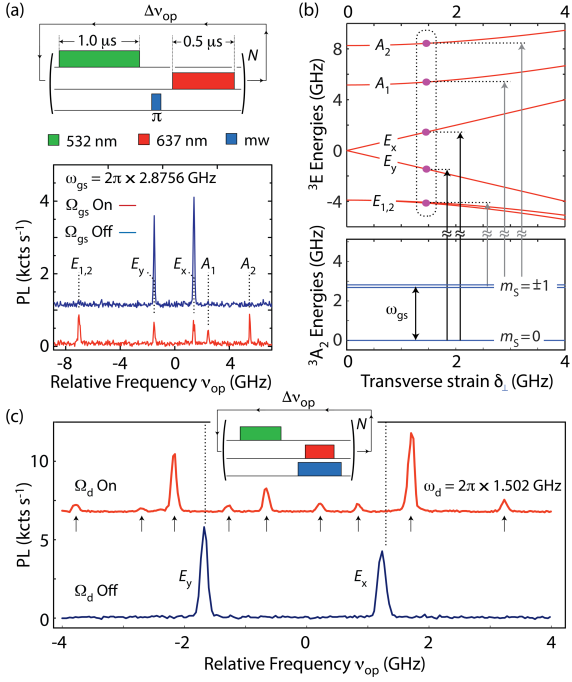

**FIG. 1: Optical spectroscopy in the presence of a mw drive.** (a) PLE spectroscopy of an individual NV center with initialization into $m_S = 0$ or $m_S = \pm 1$ in the ground state (red and blue traces, respectively). The inversion mw pulse is resonant with the ground state zero field splitting, i.e., $\omega_d = \omega_{gs}$. Here, $N = 10^5$ and $\Delta\nu_{op} = 53$ MHz. (b) Excited state $^3E$ and ground state $^3A_2$ energy diagrams (top and bottom, respectively) as a function of transverse strain. Arrows indicate allowed optical transitions from $m_S = 0$ or $m_S = \pm 1$ in the ground state (black and gray, respectively). Circles denote the measured excited state energies for the NV in (a). (c) Example PLE spectra ($N = 4 \times 10^5$, $\Delta\nu_{op} = 27$ MHz) with and without mw drive at $\omega_d = 2\pi \times 1.502$ GHz (red and blue traces, respectively). The presence of mw leads to new optical resonances (arrows). In (a) and (c), green (532 nm) and red (637 nm) blocks denote laser excitation with duration of 1 and 0.5 µs and power of 1.6 mW and 100 nW, respectively; blue squares denote mw excitation at frequency $\omega_d$. In all experiments, the temperature is 7 K; unless indicated, the laser reference frequency is 470.465 THz.

First, the displacement of the main resonance peaks away from each other hints at a coupling term shifting the energies of the mw-dressed states. We confirm this hypothesis by setting the mw tone on resonance with the $E_x$ to $E_y$ transition: Figure 2a displays PLE spectra for increasing mw power and shows for both peaks the emergence of Autler-Townes (or Rabi) splitting proportional to the amplitude of the mw drive. Indeed, in the interaction picture with $\mathcal{H}_0 = \omega_d/2(|E_x\rangle\langle E_x| - |E_y\rangle\langle E_y|)$, a coupling term $\mathcal{H}_d = \Omega_d(|E_x\rangle\langle E_y| + |E_y\rangle\langle E_x|)\sin(\omega_d t)$ leads to new eigenstates $|\pm\rangle$ in the $|E_x\rangle$, $|E_x\rangle$ basis subspace with eigenenergies $\omega_\pm = \pm\sqrt{\Omega_d + \Delta^2}/2$, where $\Delta = \omega_d - (\omega_x - \omega_y)$ is the drive detuning. Note that the presence of the two pairs of resonances at $\omega_\pm^x$ and $\omega_\pm^y$ straightforwardly arises upon introducing the laser field $H_L = (\Omega_y|0\rangle\langle E_y| + \Omega_x|0\rangle\langle E_x|)e^{\omega_L t} + c.c.$ and transforming the resulting Hamiltonian to the interaction picture[22].

Interestingly, we find that the $E_x$ and $E_y$ optical resonances are systematically sharper under a resonant mw drive. In Fig. 2c, we confirm this effect by lowering the laser power well below the saturation power to observe the inhomogeneous linewidth (broadened by spectral diffusion)[22]. This linewidth then decreases from 98 to 62 MHz, an improvement by a factor of 1.6; this effect ensues from a protection of the dressed states against electric fluctuations in the NV transverse plane. Specifically, if $\omega_x$ and $\omega_y$ denote the average frequencies of the optical resonances, a small perturbation of the transverse electric field $\epsilon_\perp$ shifts these values to $\omega_x + \epsilon_\perp$ and $\omega_y - \epsilon_\perp$. However, under a drive set at $\omega_d = \omega_x - \omega_y$ we find that a detuning $\Delta = -2\epsilon_\perp$ on the eigenenergies of the modified dressed states $|\pm\rangle'$ only shifts the optical resonances by $\pm\epsilon_\perp^2/2\Omega_d$ (to first order in $\epsilon_\perp/\Omega_d$)[22]. Intuitively, we see that for a strong enough drive, the two pairs of peaks of the dressed state spectrum remain separated by the drive frequency instead of the fluctuating splitting between the bare $E_x$ and $E_y$ lines.

Given the non-magnetic nature of these two orbitals, the coupling term likely arises from an electric transition dipole moment $\mu_{xy}$, with $\Omega_d = \boldsymbol{\mu_{xy}} \cdot \boldsymbol{E_d}$ where $\boldsymbol{E_d}$ is the electric component of the mw field. To exclude the possibility of a magnetic transition, we compare the Rabi splitting in the optical resonances with the Rabi frequency between the $m_s = 0$ and $m_s = 1$ states of the ground state manifold $^3A_2$. We then modify the impendance of the mw circuit connected to our antenna and observe different changes (+21(1)% and +40.8(4)% for the magnetic and electric Rabi frequencies, respectively), consistent with an expected change of the magnetic over electric ratio for the generated mw field. By combining those with simulations of our antenna and its environment, we can estimate the mw electric field and hence the electric transition dipole $\mu_{x,y}$. We find $|\mu_{x,y}| = 3.7 \pm 2$ Debye, depending on the precise orientation of the dipole[22].

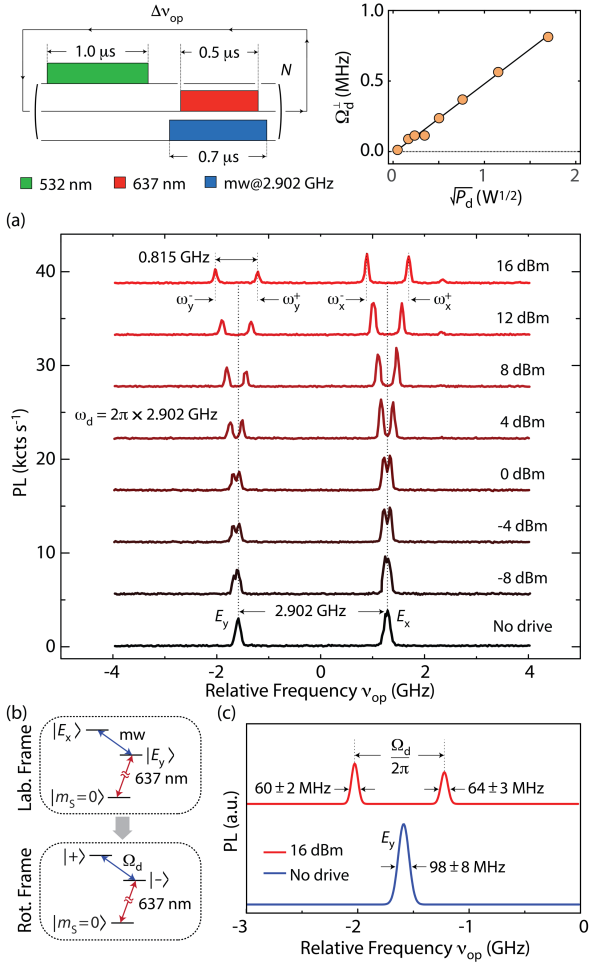

**FIG. 2: Dressed states in the $^3E$ manifold.** (a) (Main) PLE spectroscopy of a single NV under mw excitation resonant with the $|E_x\rangle \leftrightarrow |E_y\rangle$ transition for variable mw power. (Upper left) Schematics of the pulse sequence. (Upper right) Rabi splitting vs the square root of the mw power as derived from the main plot. (b) In the rotating frame description, the $|E_x\rangle$ and $|E_y\rangle$ states hybridize to yield states $|\pm\rangle$ split by the mw Rabi field $\Omega_d$. (c) Zoomed-in view of the $E_y$ optical resonance with and without mw; Rabi-split peaks show narrower linewidths.

To support these findings, we perform first principles calculations aimed at determining the amplitude and orientation of the transition dipole between the $E_x$ and $E_y$ excited states, as well as the permanent dipoles in the ground and excited states. In contrast to previous work on the dipole couplings of NV centers[28,29,30], our methodology is based on quantum embedding, treating both states of the NV$^-$ center and the dipole operator on a many-body footing[31-33]. This approach includes important (otherwise missing) electron correlations in the defect manifolds of the NV$^-$, hence leading us to expect a more accurate description of the dipole matrix elements. For our calculations, we set the strain $\sigma$ based on the susceptibility of the many-body states, such that the interpolated energy splitting between the $E_x$ and $E_y$ states approximately agrees with the experimental value[33]. We model the transition dipole between the $E_x$ and $E_y$ states and find an amplitude of 2.2 Debye, aligned with the NV crystalline axis within 10-20 degrees[22]. This value is in good agreement with the experimental estimation and is close to a similar transition dipole measured in the NV$^0$ state[34]. We also calculate the difference permanent dipole $\Delta p_x$ ($\Delta p_y$) between the $E_x$ ($E_y$) orbital and the $m_s = 0$ state, and find (in both cases) an amplitude of 2.7 Debye; further, we find that $\Delta p_x$ and $\Delta p_y$ are nearly orthogonal to each other, with a common component along the NV axis and an opposite sign component along the strain axis[22]. Our results are in good agreement with most experimental estimations[35,6,10].

These latter permanent electric dipoles can also play a role in the optical spectrum provided the external field has a component parallel to the dipole axis[6,11-13,36]: Slow changes $\delta$ of the electric field simply shift each optical resonance by $\Delta p \cdot \delta$. However, for a periodic drive whose frequency exceeds the inverse excited state lifetime — such as ours at $\omega_d$ — the Stark effect has a different impact, instead leading to a series of sidebands at multiples of $\omega_d$, also known as Landau-Zener-Stückelberg interference fringes[37-39]. The amplitude of the $n$-th sideband is typically proportional to $J_n(A/\omega_d)$, where $J_n$ is the Bessel function of the $n$-th kind and $A$ is the drive amplitude. For a weak or moderate drive, the sideband amplitudes quickly decay with increasing order: This is shown in Fig. 3a, where we measure the optical spectrum as a function of the applied mw frequency for a fixed drive amplitude, observing as expected the first two sidebands flanking the $E_y$ transition (the change in their amplitude follows the variation of the drive due to the mw circuit).

The system response is expected to change at higher amplitudes because a large enough drive should lead to a frequency comb with the main peak losing intensity. We investigate this limit in Fig. 3b for a fixed drive frequency $\omega_d = 470$ MHz; this value represents a trade-off as we try to simultaneously make $\omega_d$ greater than the inverse excited state lifetime, but sufficiently removed from the $E_x \leftrightarrow E_y$ transition frequency. Figure 3d shows the results: As the mw power increases, we first observe low-order sidebands arising more quickly around the $E_x$ line simply this results in two pairs of resonances at $\omega_\pm^x$ and $\omega_\pm^y$ with an in-pair separation equal to the generalized Rabi frequency and an out-of-pair separation equal to $\omega_d$. The sidebands generated by the Stark effect then overlap within each pair, explaining the asymmetric distribution of their amplitude. To simulate these spectra, we calculate the steady state population in the excited states, using the Lindblad master equation to describe spontaneous emission and including the electric driving of the transition and permanent dipoles as well as laser excitation from the ground state $|m_x = 0\rangle$. The right section of Fig. 3d shows the resulting spectra while Figs. 3b, 3c show the amplitude of six chosen peaks as a function of the mw amplitude. We find good aggrement with the model after performing a fit that used the amplitude of fourteen peaks while fixing parameters measured independently (such as the mw power or the mw Rabi frequency).

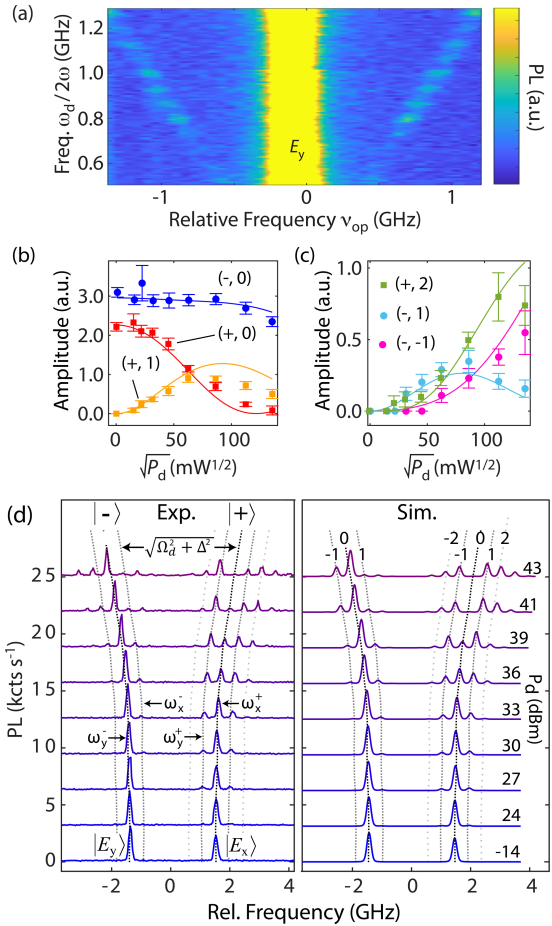

We now turn to transitions between the non-magnetic and magnetic states. Magnetic resonances in the excited states can be observed at room temperature at 1.4 GHz due to orbital averaging[16,18], but have remained elusive at low temperatures[17]. Figures 4a and 4b show continuous optically detected magnetic resonance spectra (ODMR) under green illumination (1.8 mW, above saturation) for varying mw drive power. While the regular ground state $m_s = 0$ to $m_s = \pm 1$ transition is visible at 2.87 GHz, additional features appear close to the energy difference between excited states for the present strain. On the right due to a stronger projection of the mw electric field on the $E_x$ than on the $E_y$ dipole. At higher powers a frequency comb arises as predicted but the sideband pattern is more complex than anticipated, namely, we find that (i) the central peaks corresponding to both $E_x$ and $E_y$ lines move away from each other, and (ii) the sideband amplitudes are asymmetric, contrary to what we would expect from the Bessel function symmetry. We attribute both effects to the off-resonance excitation of the $E_x \leftrightarrow E_y$ transition dipole moment: At higher powers, the Rabi frequency $\Omega_d$ becomes comparable to the detuning $\Delta = \omega_d - (\omega_x - \omega_y) \cong -2.4$ GHz, and optical resonances must consider the dressed states $|\pm\rangle$ rather than the bare states $|E_x\rangle, |E_y\rangle$. Like in Fig. 2 and as labeled for the $P_d = 33$ dBm curve, side (Fig. 4b), we observe peaks matching the $E_x \leftrightarrow A_1$ transition, while on the left side only low power measurement yields peaks close to the $E_y \leftrightarrow E_{1,2}$ transitions. In both cases, we find Lorentzian linewidths of about 58 MHz and note the absence of the $^{14}$N 40 MHz ES hyperfine splitting[40,16,41], which could be due to a hyperpolarization mechanism or to orbital mixing in the absence of external magnetic fields. At higher power, the $E_y \leftrightarrow E_{1,2}$ transitions shift to the left and broaden into multiple replicas, a feature that we now show is caused by the effect of the electric component of the mw drive.

**FIG. 3: AC Stark modulation of the NV optical transitions.** (a) PLE spectrum near the $E_y$ transition in the presence of mw of variable frequency. The reference frequency is 470.4633 THz. (b, c) Measured amplitudes of the central PLE resonance as well as four satellites as a function of $\sqrt{P_d}$. Solid lines show fits to the model. (d) PLE spectroscopy (left) and simulations (right) in the presence of an AC electric field of variable power $P_d \propto |A|^2$ and angular frequency $\omega_d = 2\pi \times 470$ MHz. As $P_d$ increases, dressed states become relevant and the main resonances show satellites separated from the center frequency by $\omega_d$. Spectra have been shifted vertically for clarity.

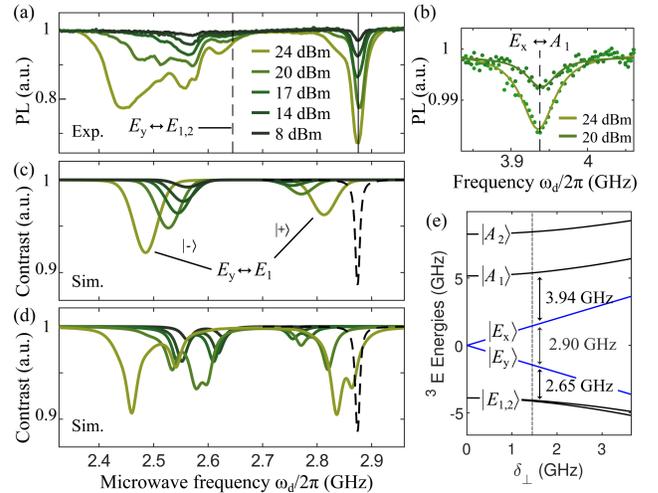

**FIG. 4: Optically detected mw resonances under green excitation.** (a, b) PL spectra at different mw powers; in (b) solid lines are Gaussian fits yielding a 58 MHz linewidth. Dashed (solid) lines are the expected position for the resonances in the excited (ground) states. Simulated ODMR spectra (c) for a single transition, considering the electric drive that splits the line in two and (d) considering the two lower transitions and accounting for mw power inhomogeneities. The dashed curve is the common $|m_S = 0\rangle \leftrightarrow |m_S = \pm 1\rangle$ ESR dip from the ground states. (e) Energy diagram and transition energies as measured from the PLE spectra.

Indeed, at frequencies close to the $E_x \leftrightarrow E_y$ transition (2.9 GHz), one must consider the dressed states rather than the bare $E_y$ orbital. Fig. 4b shows the evolution of simulated ODMR spectra with power, considering both the $E_y \leftrightarrow E_1$ transition and the $E_x \leftrightarrow E_y$ electric transition, using a simplified model[22]. Two resonances appear due to the power-dependent Rabi splitting, with the upper transition weaker due to its $E_x$ character. Crucially, this implies that the resonant frequencies depend on the mw drive power. Since the frequency response of our mw antenna is not flat, the same magnetic resonance can then be found at different frequencies. Fig. 4c shows how this affects the ODMR spectra by incorporating the frequency response of our mw circuit as measured at the antenna input[22]. Features from the experimental data (shift, broadening, resonance replica), are well-reproduced, though a quantitative match cannot be attained given the uncertainty on the exact response of our mw antenna circuit.

In conclusion, we studied the impact of mw electromagnetic fields on the short-lived NV$^-$ excited states using resonant optical excitation. We exposed the presence of a strong electric transition dipole between two of the excited-state orbitals, in addition to their permanent dipoles. With a simple wire antenna, we drove strong Rabi oscillations and created Landau-Zener-Stückelberg interference fringes in the optical spectrum. We found that dressed states are protected against fluctuations of the transverse electric field and allow tuning of the optical resonances[42] without the use of a DC electric field. First-principles calculations confirm our estimation of the transition electric dipoles and provide increased accuracy for the permanent dipoles.

The authors acknowledge helpful discussions with Mikhail Lukin. T.D. acknowledges support by the U.S. Department of Energy, Office of Science, National Quantum Information Science Research Centers, Co-design Center for Quantum Advantage (C2QA) under contract number DE-SC0012704. R.M. and C.A.M. acknowledge support from the National Science Foundation through grants NSF-2203904, and NSF-1914945. R.M. also acknowledges support from NSF-2316693. J.F. acknowledges support from grant NSF-2216838. G.I.L.M. acknowledges grant NSF-2208863. C.E.D acknowledges support from NSF grant DMR-2237674. All authors also acknowledge access to the facilities and research infrastructure of the NSF CREST IDEALS, grant number NSF-2112550.

# Supplementary Material for

# Probing electric-dipole-enabled transitions in the excited state of the nitrogen-vacancy center in diamond

Tom Delord[1*], Richard Monge[1*], Gabriel I. López-Morales[1], Olaf Bach[1], Cyrus E. Dreyer[3,4], Johannes Flick[1,2,3], and Carlos A. Meriles[1,2, †]

[1]Department. of Physics, CUNY-City College of New York, New York, NY 10031, USA.

[2]CUNY-Graduate Center, New York, NY 10016, USA.

[3]Center for Computational Quantum Physics, Flatiron Institute, New York, NY 10010, USA.

[4]Department of Physics and Astronomy, Stony Brook University, Stony Brook, New York 11794-3800, USA.

[*]Equally contributing authors

[†]Corresponding author. E-mail: cmeriles@ccny.cuny.edu

**Contents**

# I – Simulated optical spectrum via an effective model

## I.A - Hamiltonian

We construct the system Hamiltonian $\mathcal{H}$ as the sum of the energies $\mathcal{H}_0$, the effect of the laser field driving the ground to excited electric dipole $\mathcal{H}_L$, the action of the microwave (mw) field on the transition dipole moment between the $|E_x\rangle$ and $|E_y\rangle$ orbitals $\mathcal{H}_d$ (off diagonal elements of the dipole operator), as well as the interaction energy of the electric field with the permanent dipoles (diagonal elements of the dipole operator) $\mathcal{H}_P$. We limit our description to the subspace $\beta = \{|0\rangle, |E_x\rangle, |E_y\rangle\}$ spanned by $|0\rangle$, the $m_s = 0$ state in the $^3A_2$ ground state manifold, and $|E_x\rangle, |E_y\rangle$, the $m_s = 0$ states in the $^3E$ excited state manifold.

Specifically, we write

$$\mathcal{H} = \mathcal{H}_0 + \mathcal{H}_L + \mathcal{H}_d + \mathcal{H}_P, \tag{1}$$

where

$$\mathcal{H}_0 = \omega_0 |0\rangle\langle 0| + \omega_x |E_x\rangle\langle E_x| + \omega_y |E_y\rangle\langle E_y|, \tag{2}$$

$$\mathcal{H}_L = \tfrac{1}{2}(\Omega_x |0\rangle\langle E_x| + \Omega_y |0\rangle\langle E_y|) e^{i\omega_L t} + cc, \tag{3}$$

$$\mathcal{H}_d = \tfrac{1}{2}\Omega_d |E_x\rangle\langle E_y| e^{i\omega_d t} + cc, \tag{4}$$

$$\mathcal{H}_P = (A_x |E_x\rangle\langle E_x| + A_y |E_y\rangle\langle E_y|) \cos \omega_d t. \tag{5}$$

In the above formulas, we use the following notation:

- $\omega_0, \omega_x, \omega_y$ are the eigenenergies of the $|0\rangle, |E_x\rangle$ and $|E_y\rangle$ states, respectively;
- $\omega_L$ is the laser field frequency;
- $\Omega_{x,y}$ are the optical Rabi frequencies, depending on the laser intensity and orientation of the optical transition dipoles;
- $\Omega_d = \boldsymbol{\mu}_{xy} \cdot \boldsymbol{\mathcal{E}}_d$ is the Rabi frequency of the $|E_x\rangle \leftrightarrow |E_y\rangle$ drive, in turn related to $\boldsymbol{\mu}_{xy}$, the transition dipole between $|E_x\rangle$ and $|E_y\rangle$, as well as the amplitude and orientation of the mw electric field $\boldsymbol{\mathcal{E}}_d$;
- $A_{x,y} = \boldsymbol{p}_{x,y} \cdot \boldsymbol{\mathcal{E}}_d$ denote the amplitudes of the Stark effect due to the mw field, directly related to $\boldsymbol{\mathcal{E}}_d$ and the permanent electric dipoles $\boldsymbol{p}_{x,y}$ of the $|E_x\rangle$ and $|E_y\rangle$ states;
- $cc$ indicates complex conjugate.

For simplicity, we take the energy corresponding to $|0\rangle$ as the reference, and set $\omega_0 = 0$. For the Stark effect, we consider the difference of permanent dipole moments with the $|0\rangle$ state. Since the permanent dipole of a charged defect is ill-defined, we set the dipole of the $|0\rangle$ state to zero, and from now on conflate the permanent dipoles of $|E_x\rangle$ and $|E_y\rangle$ with the difference of permanent dipoles with $|0\rangle$. In all formulas, we assume $\hbar = 1$.

## I.B - Optical resonances in the interaction picture

In order to simulate the measured photoluminescence (PL) as a function of the laser field frequency, we first transform the Hamiltonian into an interaction picture by performing two unitary transformations. In the appropriate interaction picture and for a given laser frequency, we can then identify a set of optical resonances and perform a rotating wave approximation that eliminates fast-rotating terms, leading to a time-independent Hamiltonian. We then use a master equation (GKSL equation) to describe the evolution of the system under spontaneous emission, and obtain the steady state population in the optically excited states, which is proportional to the PL.

Specifically, we first perform a transformation into a mw rotating frame that eliminates the time-variation of $\mathcal{H}_d$. Let us consider:

$$\mathcal{H}_0^d = \left(\omega_y - \frac{\Delta}{2}\right)|E_y\rangle\langle E_y| + \left(\omega_x + \frac{\Delta}{2}\right)|E_x\rangle\langle E_x|, \tag{6}$$

where $\Delta = \omega_d - (\omega_x - \omega_y)$ is the detuning between the mw drive and the $|E_x\rangle \leftrightarrow |E_y\rangle$ transition frequency. Using the unitary transformation $U_d = \exp(-i\mathcal{H}_0^d t)$, the new Hamiltonian in the interaction picture $\mathcal{H}' = U_d^\dagger \mathcal{H} U_d - i U_d^\dagger dU_d/dt$ is:

$$\mathcal{H}' = \omega_+|+\rangle\langle+| + \omega_-|-\rangle\langle-| + U_d^\dagger \mathcal{H}_L U_d + U_d^\dagger \mathcal{H}_P U_d \tag{7}$$

with

$$|+\rangle = \sin\frac{\theta}{2}|E_x\rangle + \cos\frac{\theta}{2}|E_y\rangle, \tag{8}$$

$$|-\rangle = \cos\frac{\theta}{2}|E_x\rangle - \sin\frac{\theta}{2}|E_y\rangle, \tag{9}$$

$$\omega'_\pm = \pm\frac{1}{2}\sqrt{\Omega_d^2 + \Delta^2}, \tag{10}$$

$$\tan\theta = \frac{|\Omega_d|}{\Delta}. \tag{11}$$

This transformation is useful as it captures the system dynamics within the $|E_x\rangle, |E_y\rangle$ manifold in the absence of the optical drive and the Stark effect. To understand the impact of the mw drive on the optical spectrum, we look at the laser part of the Hamiltonian after the transformation, namely, $\mathcal{H}'_L = U_d^\dagger \mathcal{H}_L U_d$:

$$\mathcal{H}'_L = \frac{1}{2}e^{i\omega_L t}\left[|0\rangle\langle+|\left(\Omega_x \cos\frac{\theta}{2}e^{-i(\omega_x+\frac{\Delta}{2})t} + \Omega_y \sin\frac{\theta}{2}e^{-i(\omega_y-\frac{\Delta}{2})t}\right)\right.$$
$$\left. +|0\rangle\langle-|\left(-\Omega_x \sin\frac{\theta}{2}e^{-i(\omega_x+\frac{\Delta}{2})t} + \Omega_y \cos\frac{\theta}{2}e^{-i(\omega_y-\frac{\Delta}{2})t}\right)\right], \tag{12}$$

where $\omega_L$ denotes the laser frequency. Replacing into Eq. (7) and ignoring for now the Stark contribution, we find four optical resonances at frequencies

$$\omega^x_\pm = \frac{\omega_x + \omega_y + \omega_d}{2} + \omega_\pm, \tag{13}$$

$$\omega^y_\pm = \frac{\omega_x + \omega_y - \omega_d}{2} + \omega_\pm, \tag{14}$$

which correspond to the Rabi splittings observed in Fig. 2 of the main text. Note that the above expressions use $\Delta = \omega_d - (\omega_x - \omega_y)$ to eliminate it from the formulas. To describe the sidebands (or combs), we look at the Stark effect term after the transformation, which reads

$$U_d^\dagger \mathcal{H}_P U_d = \left[A_+|+\rangle\langle+|+A_-|-\rangle\langle-| + A_{+,-}|+\rangle\langle-|\right]\cos\omega_d t + cc, \tag{15}$$

with

$$A_+ = A_x \cos^2\frac{\theta}{2} + A_y \sin^2\frac{\theta}{2}, \tag{16}$$

$$A_- = A_x \sin^2\frac{\theta}{2} + A_y \cos^2\frac{\theta}{2}, \tag{17}$$

$$A_{+,-} = (A_y - A_x)\sin\theta/2. \tag{18}$$

The off-diagonal term is neglected, though it may have an impact at higher power (in Fig. 3, $A_{+,-}/2\omega'_+ < 0.2$ with $\sin\theta/2 < 0.38$, $A_{x,y} < 1.4$ GHz and $2\omega'_+ \cong 2.4$ GHz). To remove the time-dependence of the energy modulation, we proceed as in Ref. [1] and move into a second interaction picture using the unitary transformation:

$$U_P = \exp\left(-i\left(\omega_+ t + \frac{A_+}{\omega_d}\sin\omega_d t\right)|+\rangle\langle+| - i\left(\omega_- t + \frac{A_-}{\omega_d}\sin\omega_d t\right)|-\rangle\langle-|\right). \tag{19}$$

We can then use the Jacobi–Anger expansion and obtain:

$$U_P^\dagger |0\rangle\langle\pm| U_P = |0\rangle\langle\pm| e^{-i[\omega_\pm t + A_\pm \sin\omega_d t]}$$

$$= |0\rangle\langle\pm| \sum_{n=-\infty}^{\infty} J_n\left(\frac{A_\pm}{\omega_d}\right) e^{-i(n\omega_d + \omega_\pm)t} \tag{20}$$

where $J_n$ is the $n$-th order Bessel function. We now apply this transformation to every term in Eq. (12) and use $\Delta = \omega_d - (\omega_x - \omega_y)$ to group the terms by oscillating frequency leading to

$$U_P^\dagger U_d^\dagger \mathcal{H}_L U_d U_P = \frac{1}{2} e^{i\omega_L t} \left[|0\rangle\langle+| \sum_{n=-\infty}^{\infty} \Omega_{+,n}\, e^{-i\left(\frac{\omega_x+\omega_y+\omega_d}{2}+n\omega_d+\omega_+\right)t}\right.$$

$$\left. + |0\rangle\langle-| \sum_{n=-\infty}^{\infty} \Omega_{-,n}\, e^{-i\left(\frac{\omega_x+\omega_y+\omega_d}{2}+n\omega_d+\omega_-\right)t}\right] + cc, \tag{21}$$

with

$$\Omega_{+,n} = \Omega_x \cos\frac{\theta}{2} J_n\left(\frac{A_+}{\omega_d}\right) + \Omega_y \sin\frac{\theta}{2} J_{n+1}\left(\frac{A_+}{\omega_d}\right), \tag{22}$$

$$\Omega_{-,n} = \Omega_y \cos\frac{\theta}{2} J_n\left(\frac{A_-}{\omega_d}\right) - \Omega_x \sin\frac{\theta}{2} J_{n-1}\left(\frac{A_-}{\omega_d}\right). \tag{23}$$

In the new interaction picture, the Hamiltonian leads to a set of optical resonances at $\omega_L = (\omega_x + \omega_y + \omega_d)/2 + n\omega_d + \omega_+$ and $\omega_L = (\omega_x + \omega_y - \omega_d)/2 + n\omega_d + \omega_-$, with Rabi frequencies $\Omega_{+,n}$ and $\Omega_{-,n}$, respectively, where $n$ is an integer. In our case, weak optical excitation ensures $\Omega_{\pm,n} \ll \omega_d$ such that at a given laser frequency, we only consider the closest resonance and eliminate the others in the rotating wave approximation. We can then restrict ourselves to the two coupled states, which are described by the well-known Rabi physics[1].

**I.C – Optical lineshapes**

The optical lineshape we observe is due to a combination of homogeneous and inhomogeneous broadening. First, the homogeneous broadening can be calculated using the master equation (GKSL equation or master equation in the Lindblad form) to describe the evolution of the system under spontaneous emission and optical excitation. In our case, there are only coupling terms between two states, $|0\rangle$ and $|+\rangle$ or $|0\rangle$ and

$|-\rangle$. We assume all uncoupled states remain unpopulated and therefore consider the smaller subspace spanned by $|0\rangle$, $|e\rangle$ with $|e\rangle = |+\rangle$ or $|-\rangle$. Following the rotating wave approximation (RWA), the Hamiltonian for this simplified system becomes

$$\mathcal{H}_{RWA} = \frac{1}{2}(\Delta_R S_z + W S_x), \tag{24}$$

where $S_z = |e\rangle\langle e| - |0\rangle\langle 0|$ and $S_x = |e\rangle\langle 0| + |0\rangle\langle e|$ are the usual Paul matrices, $\Delta_R$ is the laser detuning from the resonance under consideration, and $W$ is the optical Rabi frequency.

We consider the spontaneous emission at a rate $\gamma^* = \frac{1}{2\pi \times 10.5 \text{ ns}} = 15$ MHz and write the master equation as

$$\dot{\rho} = [\rho, \mathcal{H}_{RWA}] + \gamma^* D[S_-](\rho), \tag{25}$$

where $D[S_-](\rho) = S_- \rho S_-^\dagger - \frac{1}{2}(S_-^\dagger S_- \rho + \rho S_-^\dagger S_-)$ is the Lindblad superoperator for the jump operator $S_- = |0\rangle\langle e|$, and [ ] denote commutators. We then solve the linear set of equations for all four components of the density matrix, and calculate its steady state. The population in the excited states is

$$\rho_{11} = \frac{W^2}{4\Delta_R^2 + 2W^2 + \gamma^{*2}}, \tag{26}$$

giving a Lorentzian distribution of full width at half maxima $\sqrt{2W^2 + \gamma^{*2}}$ and amplitude $\pi W/4$ (compared to a normal distribution).

To account for the inhomogeneous broadening, we simply convolute the Lorentzian with a Gaussian distribution of variance $\sigma^2$. We obtain a Voigt profile, which trends towards a Lorentzian (Gaussian) when the optical Rabi frequency $W$ is high (low) compared to the inhomogeneous broadening $\sigma$. Fits of the PLE lines under low or high optical power therefore allow us to measure both $\sigma$ and $W$ for a given power. Additionally, for a given inhomogeneous broadening, the amplitude of the PLE peak is directly related to the optical Rabi frequency. We can therefore use a peak measured at a known optical Rabi frequency $W_0$ to benchmark the relationship between the peak amplitude and $W$.

### I.D - Fit and simulated spectrum

Simple PLE spectra are fitted using Gaussian (low power) or Lorentzian (high power) distributions depending on the condition. Most spectra are obtained at 500 nW, at which a Gaussian fit works well, although the exact shape is expected to be a Voigt profile.

For more complex datasets such as in Fig. 3, we follow the procedure detailed below. The general idea is to reduce the number of parameters to fit by performing measurements of increasing complexity, starting with simpler measurements, eg in the absence of mw drive. Surprisingly, some of these parameters (e.g., optical Rabi frequency, PL rate) drift over time, in which case we use our previous measurement to initialize the fit and set bounds.

Throughout our data analysis, we first characterize the optical properties of the NV in the absence of mw. The inhomogeneous broadening $\sigma$ is determined by collecting a series of spectra under low laser power and fitting them to a Gaussian function. We find $\sigma_y = 91(5)$ MHz, $\sigma_x = 111(9)$ MHz; these values are stable in our experiments. The optical Rabi frequency $\Omega_{L,0}^{x/y}$ can then be measured for the $E_x$ or $E_y$ optical lines at a high excitation power $P_{L,0}$ by fitting the PLE spectra with a Lorentzian function or a Voigt profile. For any measurements at a laser power $P_L$, we infer the optical Rabi frequency as $\Omega_L = \sqrt{P_L/P_{L,0}} \Omega_{L,0}^h$. We find $\Omega_{L,0}^x/\sqrt{P_{L,0}} = 25.8(6)\,\text{MHz}/\sqrt{\mu W}$, $\Omega_{L,0}^y/\sqrt{P_{L,0}} = 54.9(5)\,\text{MHz}/\sqrt{\mu W}$, which leads to $\Omega_L^x = 8.16$ MHz and $\Omega_L^y = 17.4$ MHz at 100 nW of laser power. With the previous high-power measurement, we also measure a ratio in the PL amplitude with 1.6 times more PL from the $E_y$ line at equal optical Rabi frequencies, which we attribute to the different polarizations of the $E_x$ and $E_y$ lines. These parameters tend to drift over time and therefore we use these measurements as initial conditions and fit these parameters to each specific dataset.

We then characterize the strength of the mw drive of the $E_x \leftrightarrow E_y$ transition as a function of the mw power. All visible peaks (including sidebands) are fitted, and their positions are used to determine $\omega_+$ and hence $\Omega_d$. As expected, we find that $\Omega_d$ is linear with the square root of the mw power $P_d$ as measured before our cryostat input, $\Omega_d/\sqrt{P_d} = 10.7(2)$ MHz/$\sqrt{\text{mW}}$. Note that this value depends on the frequency; for 2.9 GHz we get from the fit of Fig. 2 $\Omega_d/\sqrt{P_d} = 15.8(5)$ MHz/$\sqrt{\text{mW}}$.

Finally, each peak of the data from Fig. 3 is fitted by a Gaussian and the amplitude recorded. We then compare the amplitudes as a function of the peak number and mw power to the model as described in previous sections. The plots in Fig. 3 are obtained for the following parameters: $A_x/\sqrt{P_{L,0}} = 11.7$ MHz/$\sqrt{\text{mW}}$, $A_y/\sqrt{P_{L,0}} = 19.4$ MHz/$\sqrt{\text{mW}}$, $\Omega_L^x = 11.0$ MHz, $\Omega_L^y = 6.3$ MHz and a PL ratio of 3.1. We observe a small divergence from the model for certain peaks, especially at higher powers, possibly due to drifts of the transverse field[3] or to the neglected term in Eq. (15).

## II – Simulated optically detected mw resonance spectra

To gain a qualitative understanding of the impact of the mw drive on the NV PL response under green excitation, we performed simulations using a simplified model. Here we only consider three of the excited state orbitals, $|E_x\rangle$, $|E_y\rangle$, and $|E_m\rangle$, where the latter denotes one of the magnetic states $|E_1\rangle$, $|E_2\rangle$ or $|A_1\rangle$ depending on the mw drive frequency. We consider in the following a Hamiltonian similar to the one defined in Eq. (1), namely

$$\mathcal{H}_{ES} = \mathcal{H}_0 + \mathcal{H}_d + \mathcal{H}_m, \tag{27}$$

where,

$$\mathcal{H}_0 = \omega_m |E_m\rangle\langle E_m| + \omega_x |E_x\rangle\langle E_x| + \omega_y |E_y\rangle\langle E_y|, \tag{28}$$

$$\mathcal{H}_d = \Omega_d |E_x\rangle\langle E_y| e^{i\omega_d t} + cc, \tag{29}$$

$$\mathcal{H}_m = \frac{1}{2}\Omega_m (|E_x\rangle\langle E_m| + |E_y\rangle\langle E_m|) e^{i\omega_d t} + cc, \tag{30}$$

with $\omega_m$ the energy of the $|E_m\rangle$ state[2] and $\Omega_m$ the Rabi frequency corresponding to the mw electromagnetic drive. Here, we neglect $\mathcal{H}_p$ given the high mw frequencies we work at.

We note that the Hamiltonian is the same as in Eq. (1) but substituting the $m_s = 0$ state for $|E_m\rangle$ state and taking $A_x = A_y = 0$ and $\mathcal{H}_L = \mathcal{H}_m$ with $\omega_L = \omega_d$ and $\Omega_{x,y} = \Omega_m$. We therefore perform the exact same transformation to eliminate the electric drive and find four mw resonances at the frequencies

$$\omega_{x,\pm}^m = \omega_x + \frac{\Delta}{2} \pm \frac{1}{2}\sqrt{\Omega_m^2 + \Delta^2}, \tag{31}$$

$$\omega_{y,\pm}^m = \omega_y - \frac{\Delta}{2} \pm \frac{1}{2}\sqrt{\Omega_m^2 + \Delta^2}, \tag{32}$$

with $\Delta = \omega_d - (\omega_x - \omega_y)$ and with Rabi frequencies as in Eq. (12).

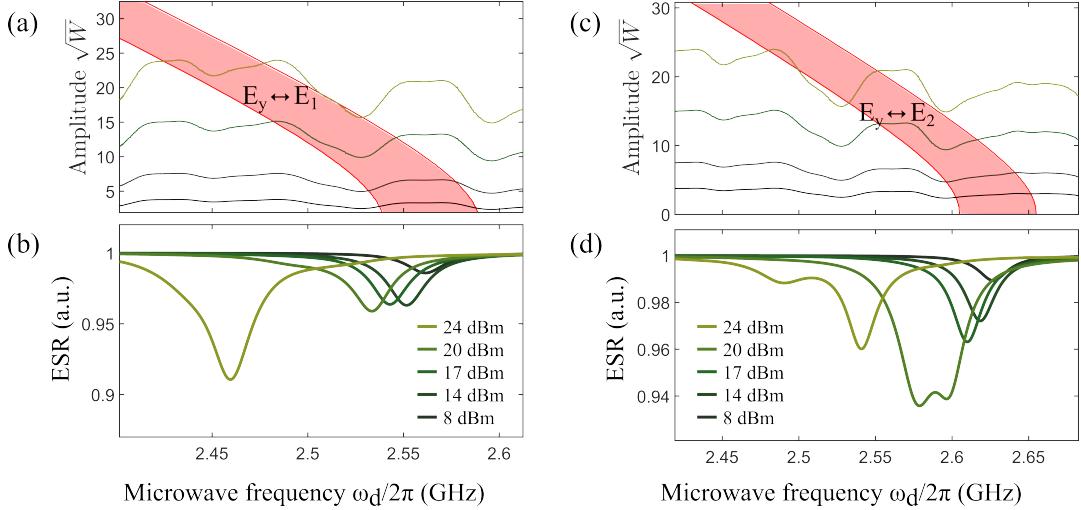

**Fig. S1: Impact of power fluctuations.** (a) Frequency response of the mw drive measured at the cryostation input for 4 different powers. Highlighted in red is the power necessary to reach resonance as a function of the frequency for the Ey to E1 transition, the width is 50 MHz (the full width at half maxima for peaks at low power). (c,d) Same but for the $E_y$ to $E_2$ transition.

We then use the formula obtained in Section I.C to estimate the population transfer out of the brighter $|E_x\rangle$ and $|E_y\rangle$ states. We take a decay rate of $\gamma^* = 2.3$ MHz (corresponding to the spin repolarization rate under green illumination) and use Rabi frequencies $\Omega_m$ estimated from the mw power and Rabi oscillations measured in the ground state under a small magnetic field ($\Omega_m/\sqrt{P_d} = 0.28$ MHz$\sqrt{\text{mW}}$ ). The electric Rabi frequency is estimated from fits of previous sections. Finally, we add contributions from different resonances and convolute the result with a 48 MHz Gaussian, which is the expected inhomogeneous broadening of the transition due to the transverse electric noise (the longitudinal part nearly cancels out). Note that a more thorough model could be obtained by solving the master equation while including the $m_s = 0$ and $m_s = 1$ ground states, one metastable singlet state, off resonant excitation, and spontaneous decay. Still, the simplified model we use is sufficient to qualitatively describe the non-trivial evolution of the spectra with the drive power. We note, however, that the simulated spectra do not capture the exact frequency response of our antenna (which likely impacts the exact position and evolution of the resonances) and do not describe the contrast well, which we leave in arbitrary units (not units of 1).

For Fig. 4c, we consider the mw power $P_d$ constant, and for Fig. 4d we use the power spectrum measured at the input of our cryostation (leading to the mw antenna) to correct $P_d$ as a function of the drive frequency $\omega_d$. Figure S1 shows the (simulated) impact the power variation has on the $|E_y\rangle \leftrightarrow |E_1\rangle$ and $|E_y\rangle \leftrightarrow |E_2\rangle$ resonances (Figs. S1b and S1d, respectively). For certain powers, we see the appearance of multiple resonances, which coincides with the presence of multiple matching conditions as shown in Figs. S1a and S1b.

## II – Set-up and characterization of the mw drive

Extensive details about the optical setup in use are given in Refs. [3-5]. Here, a more thorough characterization of the mw excitation was performed to account for non-linearities of the amplifiers. Indeed, we observed both a power saturation and the appearance of harmonics at multiple of $\omega_d$ at high mw power or on the edge of our amplifier bandwidth. Those effects need to be controlled or suppressed, especially as a single frequency excitation already leads to a complex PLE spectrum with multiple sidebands (such as Figs. 1 and 3 in the main text). To that purpose, the signal that we drive into the delivery circuit of our cryostation was analyzed by a Hewlett Packard S-Parameter Network Analyzer Model 8719ES. Throughout the manuscript, we used the power hence measured at $\omega_d$ instead of the input power fed to the amplifier. Following the observation of non-negligible harmonics at multiples of drive frequency $\omega_d$, we also used a narrow-band 380-460 MHz RF circulator to attenuate these harmonics for experiments involving 470 MHz excitation.

Microwave excitation onto the NV center is delivered by a 16-mm-long, 25-micron-wide copper wire (Thermo Fisher Scientific) overlaid over the diamond and soldered on both ends to a PCB coplanar waveguide (ExpressPCB, 50Ω). The distance between the wire and the diamond is observed under a microscope from the side and is typically under 5 µm at the center of the diamond. The mw signal is provided by a Rohde & Schwartz SMB100A signal generator and gated by a switch. For mw frequencies above 1 GHz, the signal is amplified by a Mini-circuits ZHL-16W-43-S+ 30W amplifier. For 470 MHz mw excitation, we used a Mini-circuits ZHL-30W-252-S+.

We terminate the microwave circuit with by a 50Ω resistance outside the cryostation. To confirm the electric nature of the observed transition, we momentarily remove the resistance to compare the system response with an open or closed circuit. We find that the spin Rabi frequencies in the ground state are 6.7(0.5) MHz (closed load) and 8.1(0.5) MHz (open load) or a 21(1)% increase. For the same microwave power, we measure Rabi splittings in the PLE spectra of 273.2(7) MHz (closed load) and 384.7(5) MHz (open load) corresponding to a 40.8(4)% increase. Given that both measurements are performed at almost the same mw frequency (50 MHz apart), we conclude that the spin and orbital transitions are not affected by the same component of the field, with the increased reflection under open load condition changing the electric over magnetic ratio at the NV center site.

### III – Experimental estimation of the transition dipole moment

To experimentally measure the electric transition dipole moment between the $E_x$ and $E_y$ orbitals $\boldsymbol{\mu_{xy}}$, we combine experimental measurements and simulations of the electromagnetic field generated by our antenna. We first measure $\Omega_d = \boldsymbol{\mu_{xy}} \cdot \boldsymbol{E_d}$ as a function of the drive power $P_d$, as shown in Fig. 2 of the main text. We then determine the amplitude of the microwave magnetic field at the same drive frequency $\omega_d$ in the plane orthogonal to the NV axis $B_\perp$ by performing Rabi oscillations between the $m_s = 0$ and $m_s = 1$ states in the $^3A_2$ optical ground state. Finally, we use COMSOL simulations of our antenna and its environment to estimate the ratio between the mw magnetic and electric fields at the NV position. The main source of uncertainty is the lack of knowledge on the relative orientation of the fields, of the NV, and of the transition dipole. To mitigate the uncertainty, we determine the NV orientation relative to the laboratory frame using ODMR and a magnet (roughly 10 deg. precision).

At a mw drive power of 30 dBm, we measure a magnetic Rabi frequency of 9.1 MHz, which corresponds to $B_\perp = 325$ µT. According to our COMSOL simulation and to the measured NV orientation, this leads to $|\boldsymbol{B}| = 342 \pm 26$ µT and to an electric field along the NV axis of $25 \pm 9$ kV.m$^{-1}$. Once we consider a transition dipole tilted 25 deg. from the NV axis and a random distribution for the azimuth (see next section regarding the orientation uncertainty), we find an electric field along the transition dipole of $30 \pm 13$ kV.m$^{-1}$. With a measured electric Rabi splitting of 557 MHz at this power, this leads to an estimated transition dipole moment of $3.7 \pm 2$ Debye. Note that a large uncertainty comes from possible — though unlikely — configurations where the electric field is close to orthogonal to the transition dipole. This uncertainty could be reduced by averaging over multiple NVs with different orientations.

### IV – Ab-initio calculations

All density functional theory (DFT) calculations are performed using the VASP code[6,7] with the HSE06 exchange-correlation functional[8]. We follow the quantum embedding methodology outlined in Ref. [9] to calculate the electronic states and dipole couplings of the NV on a many body footing. A 3×3×3 diamond supercell hosting a single NV center along the ⟨1, 1, 1⟩ direction is used in all calculations, with sampling of the Brillouin zone at the Gamma-point only. A plane-wave basis of 500 eV is used, with stopping criteria for electronic loops and atomic forces of $10^{-8}$ eV and $10^{-3}$ eV/Å, respectively. We consider strain along an arbitrary perpendicular direction to the ⟨1, 1, 1⟩, e.g., ⟨-1, 0, 1⟩, by deforming the supercell vectors accordingly. The strain magnitude corresponds to $2\times10^{-2}$ %, which is obtained from the strain susceptibility of the NV and a splitting that is within experimental ranges (e.g., 10 GHz). At this magnitude of strain, the calculated energy splitting between the $E_x$ and $E_y$ comes to be 30.9 GHz. This larger splitting is likely due to the absence of other interaction terms in our Hamiltonian (such as zero-flied splitting) that are likely to counteract the effects of strain but that are disregarded for simplicity. To obtain the screened Coulomb interactions between the correlated states of the NV, we employ the constrained random-phase approximation (cRPA), in which we consider ~5 empty states per atom to ensure converged results. The ground-state and strained relaxed atomic structures of the NV center are obtained using spin-polarization. Afterwards, spin-spin interactions are switched off (non-magnetic calculations), such that both spin channels can be treated on the same footing within the configuration-interaction-cRPA calculations. The same Wannierization procedure outlined in Ref. [9] is applied herein to obtain the correlated basis used in the quantum embedding calculations. Lastly, we use the matrix elements of the position operator in the Wannier basis to construct the many-body dipole operator[10]. All results presented herein are derived including the Hartree-Fock expression of the double-counting correction (see Ref. [9]), which provides reliable excitation energies for the NV center.

| | Dipole couplings: all quantities are in units of Debye (D). | | |
|---|---|---|---|
| Strain σ (×$10^{-2}$ %) | $\langle ^3E_y|\mathbf{p}|^3E_y\rangle - \langle ^3A_2|\mathbf{p}|^3A_2\rangle$ | $\langle ^3E_x|\mathbf{p}|^3E_x\rangle - \langle ^3A_2|\mathbf{p}|^3A_2\rangle$ | $\langle ^3E_x|\boldsymbol{\mu}|^3E_y\rangle$ |
| -2.00 | 0.34, -2.09, 1.63 | -0.34, 2.09, 1.63 | -0.19, -0.89, 1.99 |
| 0.00 | -1.59, 0.90, 1.63 | 1.59, -0.90, 1.63 | -0.46, 0.27, 2.38 |
| 2.00 | 0.23, -2.10, 1.63 | -0.23, 2.10, 1.63 | -0.13, -0.93, 1.99 |

**Table 1.** Components (x, y, z) of the NV⁻ dipole couplings (first two columns represent changes in permanent dipole; third column contains the transition dipoles), as derived from quantum embedding. Dipoles are expressed in the basis {⟨-1, 0, 1⟩, ⟨-1, 2, -1⟩, ⟨1, 1, 1⟩} (i.e., 'x' is strain, 'z' is the ⟨1, 1, 1⟩ crystalline axis). Strain is along ⟨-1, 0, 1⟩.

Note that the orientation of the transition dipoles between $E_x$ and $E_y$ may change with a different gauge choice of the Wannier functions. This is evidenced by performing a slight change on the Wannier basis (i.e., less localization steps), which translates into changes of < 20 deg on the net orientation of the transition dipole (the permanent dipoles remain unaffected, as they are gauge-independent quantities). Experimentally, this uncertainty is included by considering a dipole tilted 25 deg from the NV axis and a random distribution for the azimuthal angle. In any case, the behavior of the transition dipole directions under strain is that which would be expected from symmetry considerations of the system. Once the transverse strain is applied the point group symmetry is reduced from $C_{3v}$ to $C_s$, and the states with $E$ symmetry split into $A'$ and $A''$. Thus, we expect the transition dipole moment to be suppressed in the direction perpendicular to the remaining mirror plane. Indeed, in the fourth column of Table 1, we see that the moment in the direction of the strain is suppressed with respect to the unstrained case.

The results obtained are presented in Table 1. We find that the differences of permanent dipoles between the $^3A_2$ ground state and the $^3E_{x,y}$ excited states $\boldsymbol{\Delta p}_x$ and $\boldsymbol{\Delta p}_y$ have an amplitude of 2.45 Debye (no strain) and 2.67 Debye (strain) and form an angle of 97 deg (no strain) and 105 deg (strain), with $\boldsymbol{\Delta p}_x$, $\boldsymbol{\Delta p}_y$ and the NV crystalline axis in the same plane. For the transition dipole moments, we find magnitudes of 2.2 and 2.4 Debye with and without applied strain respectively, with its orientation deviating from the NV crystalline axis by 13 deg (no strain) and 25 deg (strain).

The values for the permanent dipole differences are more readily compared to experimental estimation by looking at the moments along the NV axis and perpendicular to it, as listed in Table 2. We find excellent agreement with Refs. [11,12,13] and a significant discrepancy with Refs. [14,15]. Note that the methods used differ substantially: Tamarat et al observes Stark shifts in PLE spectra of single NVs under an external electric field[12], Block et al compares the ODMR and optical spectra in NV ensembles[11] and Ji et al combines super-localization and a PLE-based measurement of NV-NV Coulomb interaction[3,14]. Likewise, from the ab initio point of view, the estimations in Ref. [15] are based on the theory of polarization within DFT, while that in Ref [13] is based on explicit electric-field Stark shifts of the excited state within DFT. Finally, the value we calculate for the transition dipole is within the confidence interval of our experimental estimation, and close to the value recently obtained for a similar transition dipole of the $NV^0$ charge state[16].

| Source | $\Delta p_\parallel$ (Debye) | $\Delta p_\perp$ (Debye) | Method |
|---|---|---|---|
| Block et al[11] | 1.4 (2) | 2.1 (2) | Optical vs ODMR |
| Tamarat et al[12] | 1.3 | – | PLE ext. field |
| Ji et al[14] | 2.82 | 3.64 | PLE triangulation |
| Alaerts et al[15] | 2.23-2.68 | – | *Ab initio* (DFT) |
| Gali[13] | 4.33 | – | *Ab initio* (DFT) |
| Present work | 1.63 | 2.12 | *Ab initio* (embedding) |

**Table 2**. Difference of permanent moments between ground and excited states along ($\Delta p_\parallel$) and perpendicular ($\Delta p_\perp$) to the NV crystalline axis.